\documentclass[12pt]{article}  

\def\[{\left\lbrack}
\def\]{\right\rbrack}

\def\({\left(}
\def\){\right)}
\newcommand{\be}{\begin{equation}}
\newcommand{\ee}{\end{equation}}
\newcommand{\ea}{\end{eqnarray}}
\newcommand{\ba}{\begin{eqnarray}}

\newcommand{\ep}{{\epsilon}}

\usepackage{graphicx}

\begin{document}

\title{Complete noncommutativity in a cosmological model with radiation}

\author{G. Oliveira-Neto and L. Fazza Marcon\\
Departamento de F\'{\i}sica, \\
Instituto de Ci\^{e}ncias Exatas, \\ 
Universidade Federal de Juiz de Fora,\\
CEP 36036-330 - Juiz de Fora, MG, Brazil.\\
gilneto@fisica.ufjf.br, luccafazzam@gmail.com}

\maketitle

\begin{abstract}
In order to try explaining the present accelerated expansion of the universe,
we consider the most complete noncommutativity, of a certain type, in a Friedmann-Robertson-Walker 
cosmological model, coupled to a perfect fluid. We use the ADM formalism in order to write the
gravitational Hamiltonian of the model and the Schutz's formalism in order to write the perfect 
fluid Hamiltonian. The noncommutativity is introduced by four nontrivial Poisson brackets
between all geometrical as well as matter variables of the model. Each nontrivial Poisson bracket
is associated to a noncommutative parameter. We recover the description in terms of commutative 
variables by introducing four variables transformations that depend on the noncommutative parameters.
Using those variables transformations, we rewrite the total noncommutative Hamiltonian of the model 
in terms of commutative variables. From the resulting Hamiltonian, we obtain the scale factor dynamical 
equations for a generic perfect fluid. In order to solve these equations, we restrict our attention to
a model where the perfect fluid is radiation. The solutions depend on six parameters: the four noncommutative 
parameters, a parameter associated with the fluid energy $C$, and the curvature parameter $k$. They also 
depend on the initial conditions of the model variables. We compare the noncommutative solutions to the
corresponding commutative ones and determine how the former ones differ from the latter ones. The comparison 
shows that the noncommutative model is very useful for describing the accelerated expansion of the universe. 
We also obtain estimates for one of the noncommutative parameters.
\end{abstract}


\section{Introduction}
\label{sec:intro}

Many years ago, in 1947, H. S. Snyder introduced noncommutative (NC) ideas in physics \cite{snyder,snyder1}. His main motivation
was the elimination of divergences in quantum field theories, by imposing noncommutativity (NCTY) between spacetime coordinates.
Since then, noncommutativity was applied to many areas of physics. In particular, noncommutativity has been extensively used 
in cosmology. As an example, some authors consider noncommutative spacetimes in order
to obtain an anisotropic dipolar imaginary primordial power spectrum. Such that, the precise power spectrum depends 
on the noncommutative geometry chosen \cite{jain,jain1}. Other authors try to constrain the energy scale of spacetime 
noncommutativity using cosmic microwave background data from the PLANCK satellite \cite{joby}. In another line of
research, some authors try to explain the accelerated expansion of the Universe using noncommutativity \cite{gil,sabido,gil1,pourhassan}.
The derivation of noncommutative cosmological models by the application of the Faddeev-Jackiw (FJ) symplectic formalism 
\cite{jackiw,barcelos,barcelos1}, was explored by other authors \cite{gil2,gil3,gil4}. Some authors have formulated
noncommutative quantum cosmology models \cite{garcia,nelson,barbosa,sabido1,gil5}. In other papers, the authors consider 
noncommutativity in order to obtain inflationary universes \cite{moniz,saba}. An interesting idea, pursued by other authors, is
how noncommutativity may induce inhomogeneity in Friedmann-Robertson-Walker (FRW) cosmology, which can be very important to the
issue of structure formation \cite{ghosh,ghosh1}.

One of the most important discoveries of the last century, in cosmology, is that the Universe is expanding in an accelerated rate \cite{riess0,perlmutter}. It means that there must be an unknown mechanism, which is opposed to gravity, acting in the
large scale Universe. That mechanism may be of a geometrical nature, in the sense that general relativity should be modified.
It can no longer be considered as the fundamental theory of gravity. Many different ideas were introduced, in the literature,
taking as the basic premise modifications of general relativity. For a review on some of those ideas see Ref. \cite{trodden}. A different line of thought considers that general relativity is the correct theory to describe gravity. Then, that unknown mechanism that drives the accelerated expansion of our Universe must be due to some, yet undiscovered, {\it exotic} type of matter. Today, in the literature, there are many different candidates to play the role of {\it exotic} matter.
For a review on some of those candidates see Ref. \cite{Mli}.

In the present work, we consider that the unknown mechanism responsible for the present accelerated expansion of the Universe is of geometrical nature. More precisely, the accelerated expansion is caused by a residual NCTY between the metric and matter variables of the model. If that NCTY really exists in nature, it must had been strong at the beginning of the Universe. Today, only a very small fraction of that NCTY must had survived. We study a model with a homogeneous and isotropic Friedmann-Robertson-Walker (FRW) metric which may have all types of constant curvatures of the spatial sections. The matter content of the model is described, initially, by a generic perfect fluid.
Later, in order to solve the dynamical equations, we restrict our attention to a model where the matter content is radiation. When we write the total commutative Hamiltonian, of the present model, we notice that it is given in terms of the following canonical variables: $(a, T, P_a, P_T)$, where $a$ is the scale factor, $T$ is a variable associated to the perfect fluid and $P_a$ and $P_T$ are, respectively, theirs canonically conjugated momenta. The NCTY considered here is not the one between spacetime coordinates first introduced by H. S. Snyder \cite{snyder,snyder1}, rather it is obtained when one imposes all possible deformed Poisson brackets between the canonical variables: $(a, T, P_a, P_T)$. As we shall see, the NCTY introduced in this way, in the present model, generates four NC parameters. That type of NCTY has already been studied in cosmology at the classical level in Refs. \cite{gil,sabido,gil1,pourhassan} and at the quantum level in Refs. \cite{garcia,nelson,barbosa,sabido1,gil5}.
The present work may be interpreted as a generalization of a previous work \cite{gil}. There, one of us, considered the same model, with a radiation perfect fluid, but the NCTY was restricted to deformed Poisson brackets between $a$ and $P_T$ $(\{a, P_T\})$ and $T$ and $P_a$ $(\{T, P_a\})$. In Ref. \cite{gil}, the NC model had just one NC parameter. In order to study the present NC model,
we start introducing its total NC Hamiltonian, with the aid of the deformed Poisson brackets between the canonical variables. Then, after using four canonical transformations, we rewrite the total NC Hamiltonian in terms of commutative variables and four NC parameters. From that total NC Hamiltonian, we obtain the scale factor dynamical 
equations for a generic perfect fluid. In order to solve these equations, we restrict our attention to
a model where the perfect fluid is radiation. We solve these equation in order to find the NC scale factor dynamics. The solutions depend on six parameters: the four noncommutative 
parameters, a parameter associated with the fluid energy $C$, and the curvature parameter $k$. They also 
depend on the initial conditions of the model variables.
Since, our main motivation is trying to interpret the accelerated expansion of the Universe, as an effect of the NCTY, we restrict our attention to expansive solutions. We compare the dynamics of both commutative and NC scale factors. Finally, using some cosmological data, we estimate the value of one NC parameter. The values we obtain are very small, in agreement with the idea that, if NCTY is really present in nature, it must be only a small residue nowadays.

In Section \ref{sec:general}, we introduce the total NC Hamiltonian of the model for a generic
perfect fluid and derive the coupled system of differential equations for the variables. In
Section \ref{sec:radiation}, we restrict our attention to the case of a radiation perfect fluid. 
We solve the system of differential equations and obtain the NC scale factor as a function
of the time coordinate and six parameters: four NC parameters, the radiation energy density $C$ and the curvature parameter
$k$. We vary all six parameters and obtain all possible expansive solutions. We compare those solutions to the ones coming 
from the corresponding commutative models. In Section \ref{sec:endofuniverse}, we use some cosmological data in order to
estimate one NC parameter. Finally, in Section \ref{sec:conclusions}, we give our conclusions.

\section{The noncommutative model for a generic perfect fluid}
\label{sec:general}

We consider that the Universe is spatially homogeneous and isotropic, therefore the metric is the FRW one, which has the following line element,
\begin{equation}  
\label{1}
ds^2 = - N^2(t) dt^2 + a^2(t)\left( \frac{dr^2}{1 - kr^2} + r^2 d\Omega^2
\right)\, ,
\end{equation}
where $a(t)$ is the scale factor, $d\Omega^2$ is the line element of the two-dimensional sphere with
unitary radius, $N(t)$ is the lapse function and $k$ is the curvature parameter. $k$ gives the type of constant curvature
of the spatial hypersurfaces. It may assume the values $k=-1$ (negative curvature), $k=1$ (positive curvature), $k=0$ (zero curvature). We are using the unit system where $c=8\pi G=1$. 

Following Weyl's postulate, we assume that the Universe is filled with a perfect fluid which has a energy-momentum tensor 
given by,
\begin{equation}
T_{\delta \nu} = (\rho+p)U_{\delta}U_{\nu} + p g_{\delta \nu}\, ,  
\label{2}
\end{equation}
where $\rho$ and $p$ are the energy density and pressure of the fluid, respectively. $U^\mu$ is the fluid four-velocity, which
in the comoving coordinate system used has the following components: $U^\mu = N(t)\delta^{\mu}_0$.
Let us impose the following equation of state for the fluid, 
\begin{equation}
\label{2.0}
p = \omega\rho,
\end{equation}
where $\omega$ is a constant which defines the type of perfect fluid.

In order to determine the perfect fluid Hamiltonian, we use the Schutz's variational formalism \cite{schutz,schutz1}. 
The main idea of this formalism is writing the four-velocity ($U_\nu$), of the fluid, in terms of six thermodynamical potentials. For the six potentials: $\mu$, $\phi$, $\zeta$, $\beta$, $\theta$ and $S$, $U_\nu$ is given by,
\be
\label{2.1}
U_\nu = \frac{1}{\mu}\left(\phi_{,\nu}+\zeta\beta_{,\nu}+\theta S_{,\nu}\right),
\ee
where $\mu$ is the specific enthalpy, $S$ is the specific entropy, $\zeta$ and $\beta$ are connected with rotation and 
$\phi$ and $\theta$ have no clear physical meaning. Since we are considering that the spatial sections of the Universe are
homogeneous and isotropic, the thermodynamical potentials $\zeta$ and $\beta$ will not contribute. We consider that the four-velocity satisfies the normalization condition,
\be
\label{2.2}
U^\nu U_\nu = -1.
\ee
Now, following the formalism, we must furnish the action ($\mathcal{S}$) for the gravity and matter sectors. For the gravity sector, the action is the Einstein-Hilbert one and for the matter sector the action is the perfect fluid one. Therefore, the combined gravity plus matter action is written as,
\be
\label{2.3}
\mathcal{S} = \int d^4x\sqrt{-g}(R + 16\pi p),
\ee
where $g$ is the determinant of the metric, $R$ is the curvature scalar and $p$ is the fluid pressure. The next step, in order
to obtain the total Hamiltonian of the model, is using the Arnowitt-Deser-Misner (ADM) formalism of general relativity and rewrite the gravitational sector of $\mathcal{S}$ \cite{misner}. Then, we introduce the metric Eq. (\ref{1}) in the resulting
expression of $\mathcal{S}$. If we also use the state of equation for the fluid, the first law of thermodynamics and some thermodynamical considerations, the resulting expression of $\mathcal{S}$ takes the form \cite{germano1},
\be
\label{2.4}
\mathcal{S}=\int dt\left[-6\frac{\dot{a}^2a}{N} + 6kNa + 
N^{-1/\omega}a^3\frac{\omega(\dot{\ep}+\theta\dot{S})^{1+1/\omega}}{(\omega+1)^{1+1/\omega}}e^{-S/\omega}\right].
\ee
Now, taking $\mathcal{S}$ Eq. (\ref{2.4}), we may obtain the Lagrangian density of the model and from it compute the total Hamiltonian of the model \cite{wheeler},
\be
\label{2.5}
N{\mathcal{H}}=N\left(-\frac{P_{a}^2}{24a} - 6ka + P_{\ep}^{\omega+1}a^{-3\omega}e^S\right),
\ee
where $P_a = -12\dot{a}a/N$ and $P_\ep = N^{-1/\omega}a^3(\dot{\ep}+\theta\dot{S})^{(\omega+1)^{-1/\omega}/\omega}e^{-S/\omega}$.
We may simplify the total Hamiltonian Eq. (\ref{2.5}), by performing the following canonical transformations \cite{rubakov},
\be
\label{2.6}
T = -P_S e^{-S}P_\ep^{-(\omega+1)},\quad P_T = P_\ep^{\omega+1}e^S,\quad \bar{\ep} = \ep-(\omega+1)\frac{P_S}{P_\ep},\quad \bar{P_\ep} = P_\ep,
\ee
where $P_S = \theta P_\ep$. Using the above transformations the total Hamiltonian Eq. (\ref{2.5}) is simplified to,
\begin{equation}
N {\mathcal{H}}= -\frac{P_{a}^2}{24} - 6ka^2 + a^{1-3\omega}P_{T},  
\label{3}
\end{equation}
where $P_{a}$ and $P_{T}$ are the momenta canonically conjugated to $a$ and 
$T$, the latter being the canonical variable associated to the fluid.
In the present work, we choose the gauge where $N = a$.

Equation (\ref{3}) gives the total Hamiltonian of the commutative model. Now, we want to write the total NC Hamiltonian of the
model. In order to do that, we consider that the total NC Hamiltonian has the same functional form of Eq. (\ref{3}). But now it is written in terms of NC variables,
\begin{equation}
N_{nc} {\mathcal{H}}_{nc}= -\frac{P_{a_{nc}}^2}{24} - 6ka_{nc}^2 + a_{nc}^{1-3\omega}P_{T_{nc}},  
\label{3,5}
\end{equation}
Next, we impose that the noncommutative variables of the model \linebreak
$\{a_{nc}, P_{a_{nc}}, T_{nc}, P_{T_{nc}}\}$ satisfy the following deformed Poisson brackets (PBs):
\ba
\label{4}
\left\{a_{nc},T_{nc}\right\} & = & \sigma,\\
\label{4.1}
\left\{P_{a_{nc}},P_{T_{nc}}\right\} & = & \alpha,\\
\label{4.2}
\left\{T_{nc},P_{a_{nc}}\right\} & = & \chi,\\
\label{4.3}
\left\{a_{nc},P_{T_{nc}}\right\} & = & \gamma.
\ea
We also impose the usual PBs,
\be
\label{4.4}
\left\{a_{nc},P_{a_{nc}}\right\}  =  \left\{T_{nc},P_{T_{nc}}\right\} = 1.
\ee 
Where $\sigma$, $\alpha$, $\chi$ and $\gamma$ are the NC parameters. It is important to notice that these are
all possible deformed PBs one may propose, for the present model. Once, we are considering that the NCTY is a small
residual effect, nowadays, the NC parameters appears to first order in the deformed PBs Eqs. (\ref{4})-(\ref{4.3}).
The present work may be interpreted as a generalization of a previous work \cite{gil}. There, one of us, considered the same model, with a radiation perfect fluid, but the NCTY was restricted to deformed PBs between $T$ and $P_a$ $(\{T, P_a\})$ and $a$ and $P_T$ $(\{a, P_T\})$. In Ref. \cite{gil}, these two deformed PBs have identical values leading to a single NC parameter. Here, besides these two deformed PBs have different values Eqs. (\ref{4.2}) and (\ref{4.3}), we demand that the other possible deformed PBs  Eqs. (\ref{4}) and (\ref{4.1}) are, also, non-trivial. Then, in the present NC model we have four NC parameters.

In order to simplify our description of the present model, we want to introduce canonical transformations connecting the NC variables: \linebreak $\{a_{nc}, P_{a_{nc}}, T_{nc}, P_{T_{nc}}\}$, with new commutative ones: $\{a_{c}, P_{a_c}, T_{c}, P_{T_c}\}$. These new
commutative variables must satisfy the usual PBs. Those type of transformations were first introduced in Refs. \cite{susskind,mezincescu,morariu} and sometimes are called Bopp shift \cite{zachos,zachos1,gamboa,kokado}.
Taking in account the deformed PBs Eqs. (\ref{4})-(\ref{4.3}), one of the most general transformations, to first order in NC parameters $\sigma$, $\alpha$, $\chi$ and $\gamma$, leading from the NC variables to new commutative ones, are given by,
\ba
\label{5}
a_{nc} & = & a_c + \frac{\gamma}{2}T_c - \frac{\sigma}{2}P_{T_c},\\
\label{5.1}
P_{a_{nc}} & = & P_{a_c} + \frac{\chi}{2}P_{T_c} + \frac{\alpha}{2}T_c,\\
\label{5.2}
T_{nc} & = & T_c + \frac{\chi}{2}a_c + \frac{\sigma}{2}P_{a_c},\\
\label{5.3}
P_{T_{nc}} & = & P_{T_c} + \frac{\gamma}{2}P_{a_c} - \frac{\alpha}{2}a_c,
\ea
If we introduce the values of the NC variables, Eqs. (\ref{5})-(\ref{5.3}), in the deformed PBs Eqs. (\ref{4})-(\ref{4.3}) and demand that the commutative variables satisfy the usual PBs, we can show that the NC variables satisfy them to first in the NC parameters. Those transformations Eqs. (\ref{5})-(\ref{5.3}) are not unique. One of the motivations for using them is that they reduce to the transformations introduced in Ref. \cite{gil}, when one sets $\sigma=\alpha=0$ and $\chi=\gamma$.
Now, we want to describe the NC scale factor, $a_{nc}$ Eq. (\ref{5}), time evolution by computing the Hamilton's equation, for
the present model. Therefore, we start rewriting the total NC Hamiltonian, $N_{nc} {\mathcal{H}}_{nc}$ Eq. (\ref{3,5}), in terms of the commutative variables Eqs. (\ref{5})-(\ref{5.3}), in the gauge $N_{nc}=a_{nc}$,
\ba
\mathcal{H}_{nc} & = & \frac{1}{12}\left(P_{a_{c}}+\frac{\chi P_{T_{c}}}{2}+\frac{\alpha T_{c}}{2}\right)^{2} - 3k\left(a_{c}+\frac{\gamma T_{c}}{2}-\frac{\sigma P_{T_c}}{2}\right)^{2}\nonumber\\
& + & \left(P_{T_{c}}+\frac{\gamma P_{a_{c}}}{2}-\frac{\alpha a_{c}}{2}\right) \left(a_{c}+\frac{\gamma T_{c}}{2}-\frac{\sigma P_{T_c}}{2}\right)^{1-3\omega}.
\label{6}
\ea
From the Hamiltonian Eq. (\ref{6}), we compute the Hamilton's equation with the aid of the usual PBs among the commutative variables. They are,
\ba
\label{7}
\dot{a}_{c} & = & \frac{\partial\mathcal{H}_{nc}}{\partial P_{a_{c}}} = \frac{1}{6}\left(P_{a_{c}}+\frac{\chi P_{T_{c}}}{2}+\frac{\alpha T_{c}}{2}\right) +\frac{\gamma}{2}\left(a_{c}+\frac{\gamma T_{c}}{2}-\frac{\sigma P_{T_{c}}}{2}\right)^{1-3\omega},\\
\label{7.1}
\dot{P}_{a_{c}} & = & - \frac{\partial\mathcal{H}_{nc}}{\partial a_{c}} = 6k\left(a_{c}+\frac{\gamma T_{c}}{2}-\frac{\sigma P_{T_{c}}}{2}\right) + \frac{\alpha}{2}\left(a_{c}+\frac{\gamma T_{c}}{2}-\frac{\sigma P_{T_{c}}}{2}\right)^{1-3\omega}\nonumber\\
& - & (1-3\omega)\left(P_{T_{c}}+\frac{\gamma P_{a_{c}}}{2}-\frac{\alpha a_{c}}{2}\right)\left(a_{c}+\frac{\gamma T_{c}}{2}-\frac{\sigma P_{T_{c}}}{2}\right)^{3\omega},\\
\label{7.2}
\dot{T}_{c} & = & \frac{\partial\mathcal{H}_{nc}}{\partial P_{T_{c}}} = \frac{\chi}{12}\left(P_{a_{c}}+\frac{\chi P_{T_{c}}}{2}+\frac{\alpha T_{c}}{2}\right) + 3k\sigma\left(a_{c}+\frac{\gamma T_{c}}{2}-\frac{\sigma P_{T_{c}}}{2}\right)\nonumber\\ 
& + & \left[1-\frac{\sigma}{2}P_{T_{c}}(1-3\omega)\left(a_{c}+\frac{\gamma T_{c}}{2}-\frac{\sigma P_{T_{c}}}{2}\right)^{-1}\right]\left(a_{c}+\frac{\gamma T_{c}}{2}-\frac{\sigma P_{T_{c}}}{2}\right)^{1-3\omega}\nonumber\\
& - & \frac{\sigma}{2}(1-3\omega)\left(\frac{\gamma P_{a_{c}}}{2}-\frac{\alpha a_{c}}{2}\right)\left(a_{c}+\frac{\gamma T_{c}}{2}-\frac{\sigma P_{T_{c}}}{2}\right)^{3\omega},\\
\label{7.3}
\dot{P}_{T_{c}} & = & - \frac{\partial\mathcal{H}_{nc}}{\partial T_{c}} = -\frac{\alpha}{12}\left(P_{a_{c}}+\frac{\chi P_{T_{c}}}{2}+\frac{\alpha T_{c}}{2}\right) + 3k\gamma\left(a_{c}+\frac{\gamma T_{c}}{2}-\frac{\sigma P_{T_{c}}}{2}\right)  \nonumber\\
& - & \frac{\gamma}{2}(1-3\omega)\left(P_{T_{c}}+\frac{\gamma P_{a_{c}}}{2}-\frac{\alpha a_{c}}{2}\right)\left(a_{c}+\frac{\gamma T_{c}}{2}-\frac{\sigma P_{T_{c}}}{2}\right)^{-3\omega}.
\ea

Now, we would like to find the NC scale factor behavior Eq. (\ref{5}). In the general situation, for generic $\omega$ and $k$, the best we can
do is writing, from Eqs. (\ref{7})-(\ref{7.3}), a system of two coupled differential equations involving $a_c(t)$, $T_c(t)$
and their time derivatives. This is done in the following way. Combining Eqs. (\ref{7}), (\ref{7.1}) and (\ref{7.3}), we obtain the following
relationship between $P_{T_c}$ and $P_{a_c}$,
\be
\label{11}
P_{T_{c}} = C-\frac{\alpha}{2}a_{c} + \frac{\gamma}{2}P_{a_{c}},
\ee
where $C$ is an integration constant. Physically, for the commutative case ($\sigma =\alpha =\chi =\gamma =0$), $C$ represents the fluid energy, which means 
that it is positive. Then, using Eqs. (\ref{7}) and (\ref{7.2}), we find, to first order in the NC parameters, the following equation
expressing $P_{a_c}$ in terms of time derivatives of $a_c$ and $T_c$,
\be
\label{12}
P_{a_{c}}= -6\dot{a_{c}} - \frac{\chi}{2}P_{T_{c}} - \frac{\alpha}{2}T_{c} + 3\gamma \dot{T_{c}}.
\ee
Now, we derive the last equation with respect to the time $t$ and find,
\begin{equation}
\label{13}
\ddot{a_{c}} = \frac{-1}{6}\left(\dot{P_{a_{c}}}+\frac{\chi}{2}\dot{P_{T_{c}}} + \frac{\alpha}{2}\dot{T_{c}}\right) + \frac{\gamma}{2}\ddot{T_{c}}.
\end{equation}
Next, we, also, derive equation (\ref{7.2}) with respect to the time $t$, in order to obtain the value of $\ddot{T_c}$. After that, we multiply $\ddot{T_c}$ by $\gamma/2$ and reach the following value for this product, to first order in the NC parameters,
\begin{equation}
\label{13.5}
\frac{\gamma}{2}\ddot{T_{c}} = \frac{\gamma}{2}(1-3\omega)\left(a_{c}+\frac{\gamma T_{c}}{2}-\frac{\sigma P_{T_{c}}}{2}\right)^{-3\omega}\left(\dot{a_{c}}+\frac{\gamma \dot{T_{c}}}{2}-\frac{\sigma \dot{P_{T_{c}}}}{2}\right).
\end{equation}
The first equation of the system, to be achieved is the one for $a_c(t)$ and its time derivatives. In order to do that,
we introduce the values of $\dot{P}_{a_c}$ Eq. (\ref{7.1}), $\ddot{T}_c$ Eq. (\ref{13.5}), $\dot{P}_{T_c}$ Eq. (\ref{7.3}), 
$P_{T_c}$ Eq. (\ref{11}) and $P_{a_c}$ Eq. (\ref{12}), in Eq. (\ref{13}). It gives, to first
order in NC parameters, the following differential equation for $a_c$,
\ba
\ddot{a_{c}} & = & -k\left(a_{c}+\frac{\gamma T_{c}}{2}-\frac{\sigma C}{2}\right)
 - \frac{1}{2}\omega(1-3\omega)C\left(\frac{\gamma}{2}T_{c}-\frac{\sigma}{2}C\right)a_{c}^{1-3\omega}\nonumber\\ 
& - &\frac{\alpha}{6}a_{c}^{1-3\omega} + \frac{1}{6}(1-3\omega)(C-3\gamma\dot{a_{c}}  - \alpha a_{c})a_{c}^{-3\omega}.
\label{14}
\ea
Finally, in order to attain the equation for $T_c(t)$ and its time derivatives, we introduce the values of $P_{T_c}$ Eq. (\ref{11}) and $P_{a_c}$ Eq. (\ref{12}), in Eq. (\ref{7.2}). To first order in the NC parameters, it leads to,
\begin{equation}
\dot{T_{c}} = \frac{\chi}{2}\dot{a_{c}} + 3k\sigma a_{c} - C\frac{\sigma}{2}(1-3\omega)a_{c}^{-3\omega} + (1-3\omega)\left(\frac{\gamma}{2}T_{c} - \frac{\sigma}{2}C\right)a_{c}^{1-3\omega}.
\label{14.5}
\end{equation}
All the information about the noncommutativity is encoded in the NC parameters. If we set them to zero we recover the
commutative model in the gauge $N=a$. In particular, equation (\ref{14}) decouples and we may solve it to obtain the scale factor dynamics. In order to solve those equations and compute $a_{nc}$ Eq. (\ref{5}), we shall have to furnish initial conditions for $a_c(t)$, $\dot a_c(t)$ and $T_c(t)$.

Another important equation is the NC Friedmann equation. We must use it in order to compute the physically acceptable initial
conditions for $\dot a_c(t)$, given the initial conditions for $a_c(t)$ and $T_c(t)$. It may be computed by imposing the superhamiltonian constraint: $\mathcal{H}_{nc}=0$. Therefore, with the aid of Eqs. (\ref{6}), (\ref{11}) and (\ref{12}), the
NC Friedmann equation is given, to first order in the NC parameters, by,
\ba
& - & 3\dot{a_{c}}^{2} - 3\gamma\dot{a_{c}} a_{c}^{1-3\omega} -3ka_{c}^{2} - 6ka_{c}\left(\frac{\gamma}{2}T_{c} - \frac{\sigma}{2}C\right)\nonumber\\ 
& + & Ca_{c}^{1-3\omega} + (1-3\omega)C\left(\frac{\gamma}{2}T_{c} - \frac{\sigma}{2}C\right)a_{c}^{1-3\omega} - \alpha a_{c}^{2-3\omega} = 0. 
\label{14.6}
\ea
If we set all the NC parameters to zero in Eq. (\ref{14.6}), we recover the commutative Friedmann equation in the gauge $N=a$.
Before trying to solve the system of differential equations (\ref{14}) and (\ref{14.5}), we may rewrite $a_{nc}$ Eq. (\ref{5}) in a simpler way, to first order in the NC parameters, using Eq. (\ref{11}),
\begin{equation}
a_{n c}(t)=a_{c}+\frac{\gamma T_{c}}{2}-\frac{\sigma C}{2}.
\label{14.7}
\end{equation}
It is important to stress that $a_{nc}(t)$, Eq. (\ref{14.7}), is the physical scale factor of the present NC model. Therefore, it is
that quantity that we shall try to evaluate and estimate, in the next Sections.

\section{The noncommutative model for a radiation perfect fluid}
\label{sec:radiation}

In order to find solutions to $a_c(t)$ and $T_c(t)$, from the system Eqs. (\ref{14}) and (\ref{14.5}), we must fix the value of the parameter $\omega$. In other words, we must choose a perfect fluid to represent the matter content of the Universe. In the present work, let us choose the radiation perfect fluid with $\omega=1/3$. That type of matter must had been important at the beginning of the Universe \cite{wheeler}. As explained in Section \ref{sec:intro}, we want to describe the present accelerated expansion of the Universe, not by the presence of a matter component, but due to the NCTY. Therefore, the present model has a matter component associated to the early Universe and NCTY to explain the present accelerated expansion of the Universe. We plan to study a more complete model of the Universe in a future work.
Therefore, if we impose that the matter content of our NC model is radiation, the system of dynamical equations Eqs. (\ref{14}) and (\ref{14.5}), simplify to,
\ba
\label{14.8}
\ddot{a}_{c} & = & -k\left(a_{c}+\frac{\gamma T_{c}}{2}-\frac{\sigma C}{2}\right)-\frac{\alpha}{6},\\
\label{14.9}
\dot{T_{c}} & = & \frac{\chi}{2}\dot{a_{c}} + 3k\sigma a_{c},
\ea
and the NC Friedmann equation (\ref{14.6}) to,
\begin{equation}
\label{14.95}
-3 \dot{a}_{c}^{2}-3 \gamma \dot{a}_{c}-3 k a_{c}^{2}-6 k a_{c}\left(\frac{\gamma T_{c}}{2}-\frac{\sigma C}{2}\right)+C-\alpha a_{c} = 0.
\end{equation}

In order to solve the system of differential equations (\ref{14.8}) and (\ref{14.9}), we believe that the best way to do that is
fixing, initially, the value of $k$, for each different curvature. Then, for each curvature, we solve the corresponding system of differential equations (\ref{14.8}) and (\ref{14.9}) and investigate how $a_{nc}(t)$ Eq. (\ref{14.7}) behaves for different values of: $\sigma$, $\alpha$, $\chi$, $\gamma$, $C$ and the initial conditions for $a_c(t)$, $\dot{a}_{c}(t)$ and $T_c(t)$. 
When we are not studying the initial conditions of $a_c(t)$ and $T_c(t)$, they will be fixed at the following values,
\begin{equation}
T_{c}(t=0) \equiv T_{0} = 0, 
\label{ic1}
\end{equation}
\begin{equation}
a_{c}(t=0) \equiv a_{0} = 1. 
\label{ic2}
\end{equation}
When we are studying those initial conditions $a_0$ and $T_0$, they may assume any positive value.
Eq. (\ref{14.8}) is a second order differential equation for $a_c(t)$. Therefore, in order to find a solution to that equation, we need two initial conditions: $a_c(t=0)=a_0$ and $\dot{a}_{c}(t=0) \equiv v_0$. For given values of $a_0$, $T_0$ and the other parameters, we obtain $v_0$ using the NC Friedmann equation (\ref{14.95}). In order to do that, we introduce the values of $a_0$, $T_0$ and the other parameters in Eq. (\ref{14.95}) and we solve the resulting algebraic equation to $v_0$. We restrict our attention to positive values of $v_0$, because we want expansive solutions.

\subsection{The case $k = 0$}
\label{k=0}

Now fixing $k=0$, the spatial sections have no curvature or, in other words, they are flat. Introducing $k=0$ in the system Eqs. (\ref{14.8}) and (\ref{14.9}) and in the NC Friedmann equation (\ref{14.95}), we obtain,
\ba
\label{18.0}
\ddot{a}_{c} & = & -\frac{\alpha}{6},\\
\label{18.1}
\dot{T_{c}} & = & \frac{\chi}{2}\dot{a_{c}},
\ea
\begin{equation}
\label{18.2}
3 \dot{a}_{c}^{2} + 3\gamma \dot{a}_{c} - C + \alpha a_{c} = 0.
\end{equation}
In the present case, as we can see from Eqs. (\ref{18.0}) and (\ref{18.1}), the system of equations decouple and we may find algebraic
solutions to $a_c$ and $T_c$. After solving the system Eqs. (\ref{18.0}) and (\ref{18.1}), taking in account the initial condition
$v_0$ coming from the NC Friedmann equation (\ref{18.2}), we obtain solutions to $a_c$ and $T_c$. Combining these solutions
according to Eq. (\ref{14.7}), we find the following physically acceptable expression for $a_{nc}$,
\begin{equation}
a_{nc}(t) = -\frac{\alpha}{12}t^{2} + \left( -\frac{\gamma}{2} + \frac{\sqrt{9\gamma^{2}+12(C - \alpha a_{0})}}{6} \right)t + a_{0} - \frac{\sigma C}{2} + \frac{\gamma}{2}T_0. 
\label{19.0}
\end{equation}
If we set all NC parameters to zero in Eq. (\ref{19.0}), we obtain the commutative scalar factor ($a(t)$),
\begin{equation}
a(t) = \sqrt{\frac{C}{3}} t + a_0.
\label{19.1}
\end{equation}
As we can see from Eq. (\ref{19.1}), $a(t)$ expands as a linear function of $t$. The expansion rate increases with the increase
of $C$.

Now, we want to investigate the behavior of $a_{nc}(t)$ Eq. (\ref{19.0}) with respect to $C$ and the different NC parameters: $\alpha$, $\gamma$, $\sigma$. As we can see, from that expression, $a_{nc}(t)$ does not depend on the NC parameter $\chi$. We, also, want to investigate how $a_{nc}(t)$ Eq. (\ref{19.0}), varies with the initial conditions $a_0$, $v_0$ and $T_0$. It is important to remember that, the values of $a_0$ and $T_0$ will be fixed following Eqs. (\ref{ic1}) and (\ref{ic2}), when we are not studying them. The values of $v_0$ have already been incorporated in $a_{nc}(t)$ Eq. (\ref{19.0}). When we are studying $v_0$, we choose different values of that initial condition and let $C$ varies. The different values of $C$ are obtained with the aid of the NC Friedmann equation (\ref{18.2}). Finally, we want to compare the NC scale factor Eq. (\ref{19.0})
with the commutative one Eq. (\ref{19.1}). In all the examples we give in the next Subsections, the values of the parameters
and initial conditions are chosen for a better visualization of the results.

\subsubsection{Varying $\alpha$}

\begin{figure}
\begin{center}
\includegraphics[height=7.5cm,width=9.5cm]{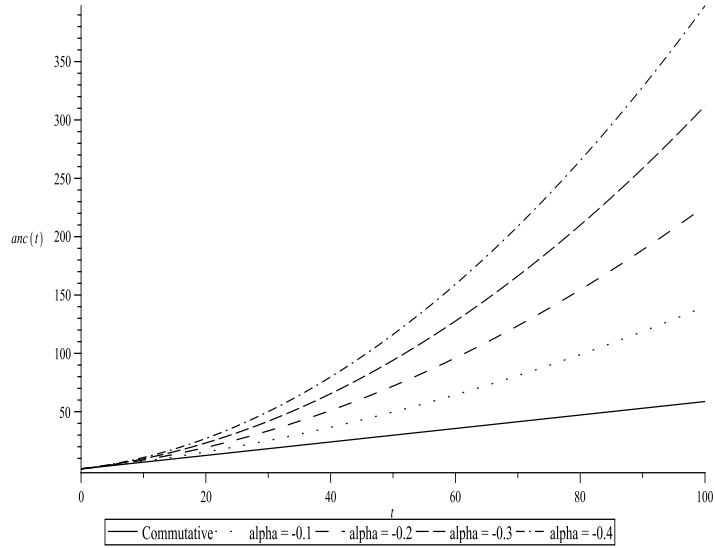}
\end{center}
\caption{$a_{nc}(t)$ as a function of $t$ for different values of $\alpha$ and $C = 1$, $\sigma = -0.1$, $\gamma = 0.1$}
\label{k0_alpha}
\end{figure}

As we can see from $a_{nc}(t)$ Eq. (\ref{19.0}), we must impose that $\alpha \leq 0$, in order to obtain expansive solutions. If we do that, the general behavior of $a_{nc}(t)$ Eq. (\ref{19.0}) describes an universe that starts
to expand, in an accelerated rate, from its initial size at $t=0$, and continues to expand to an infinity size, after
an infinite time interval. For fixed values of the other parameters and initial conditions, we observe that, if $\alpha$ diminishes, the more quickly the NC scale factor Eq. (\ref{19.0}) expands. As an example, we may see Figure \ref{k0_alpha}.

\subsubsection{Varying $\gamma$}
\label{gamma}

\begin{figure}
\begin{center}
\includegraphics[height=7.5cm,width=9.5cm]{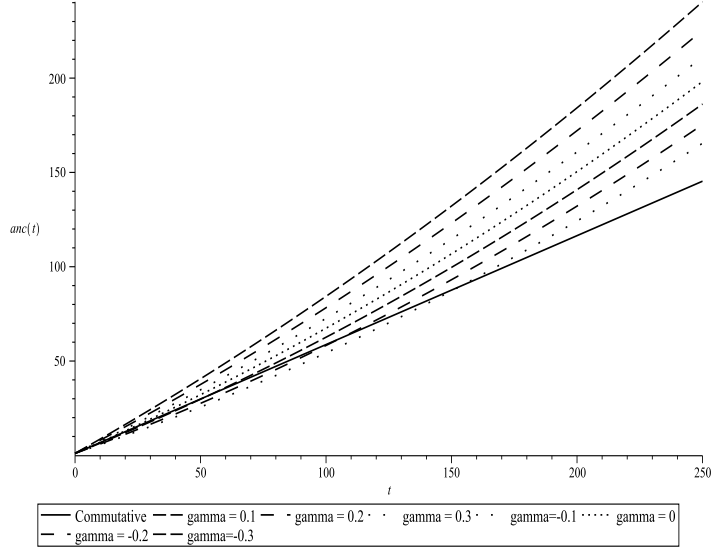}
\end{center}
\caption{$a_{nc}(t)$ as a function of $t$ for different values of $\gamma$ and $C = 1$, $\sigma = -0.1$, $\alpha = -0.01$}
\label{k0_gamma}
\end{figure}

Initially, if $T_0 \geq 0$ and $\gamma < 0$ we notice from Eq. (\ref{19.0}), that $a_{nc}(t)$ may be negative. Therefore, since $\gamma$ may be positive or negative, in general, we shall restrict our attention to the situation where $T_0 = 0$. It means that, we shall not investigate how $a_{nc}(t)$ depends on $T_0$, for the present models with $k=0$. For expansive solutions, with $\alpha \leq 0$, if we fix the values of the other parameters and initial conditions, $a_{nc}(t)$ Eq. (\ref{19.0}) expands more quickly for smaller values of $\gamma$. As an example, we may see Figure \ref{k0_gamma}.

\subsubsection{Varying $\sigma$}
\label{sigma}

\begin{figure}
\begin{center}
\includegraphics[height=7.5cm,width=9.5cm]{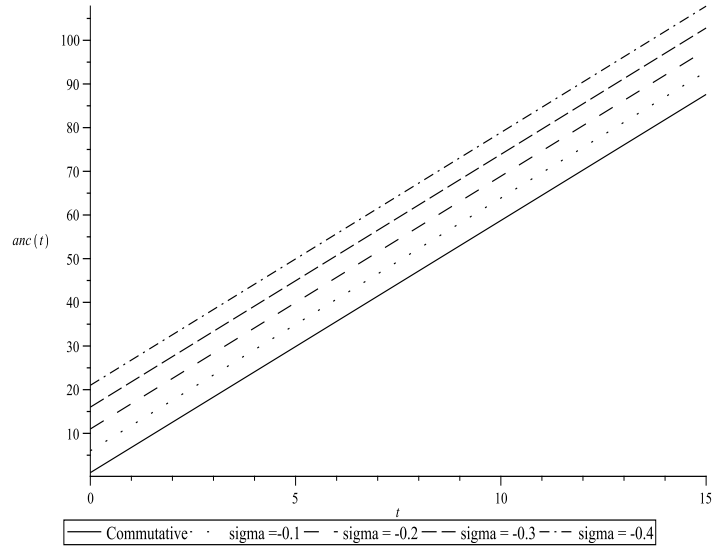}
\end{center}
\caption{$a_{nc}(t)$ as a function of $t$ for different values of $\sigma$ and $C = 100$, $\gamma = 0.1$, $\alpha = -0.05$}
\label{k0_sigma}
\end{figure}

As we can see from $a_{nc}(t)$ Eq. (\ref{19.0}), $\sigma$ gives a constant contribution to the solutions. It is important to notice that 
$\sigma$ has to be negative or zero. That is the case because $C \geq 0$ and if $\sigma > 0$, $a_{nc}(t)$ may be negative, from 
Eq. (\ref{19.0}). Only for $\sigma \leq 0$, $a_{nc}(t)$ is always positive, for any value of $C$. For smaller
values of $\sigma \leq 0$, and keeping fix the other parameters and initial conditions, the initial value of $a_{nc}(t)$ increases. After that, the solutions with different values of $\sigma$ and different initial conditions expand at the same rate. Therefore, for smaller values of $\sigma$ the solutions have greater values, for a given instant of time. As an example, we may see Figure \ref{k0_sigma}.

\subsubsection{Varying $C$}
\label{C}

\begin{figure}
\begin{center}
\includegraphics[height=7.5cm,width=9.5cm]{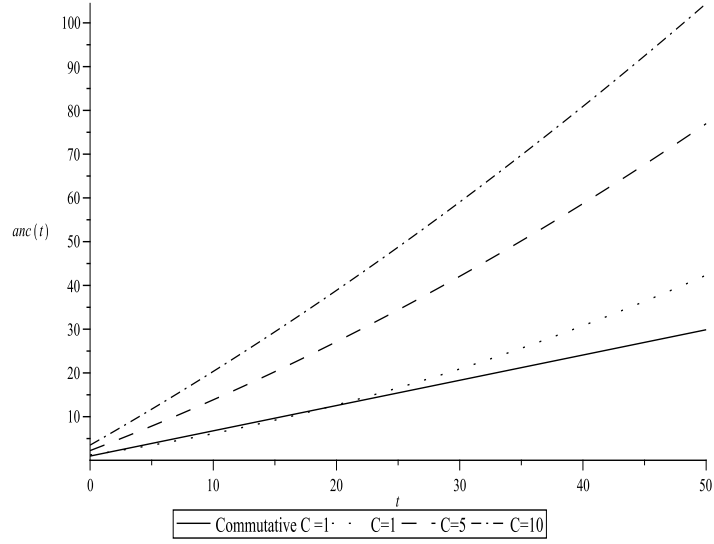}
\end{center}
\caption{$a_{nc}(t)$ as a function of $t$ for different values of $C$ and $\sigma = -0.5$, $\gamma = 0.5$, $\alpha = -0.1$}
\label{k0_C}
\end{figure}

As we can see from $a_{nc}(t)$ Eq. (\ref{19.0}), if $\alpha \leq 0$, for any positive value of $C$ the solutions are always
expansive. If we fix the values of the other parameters and initial conditions, the solutions expand more quickly for greater values of $C$ and the initial value of $a_{nc}(t)$ increases when $C$ increases. As an example, we may see Figure \ref{k0_C}.

\subsubsection{Varying $a_0$}
\label{a0}

\begin{figure}
\begin{center}
\includegraphics[height=7.5cm,width=9.5cm]{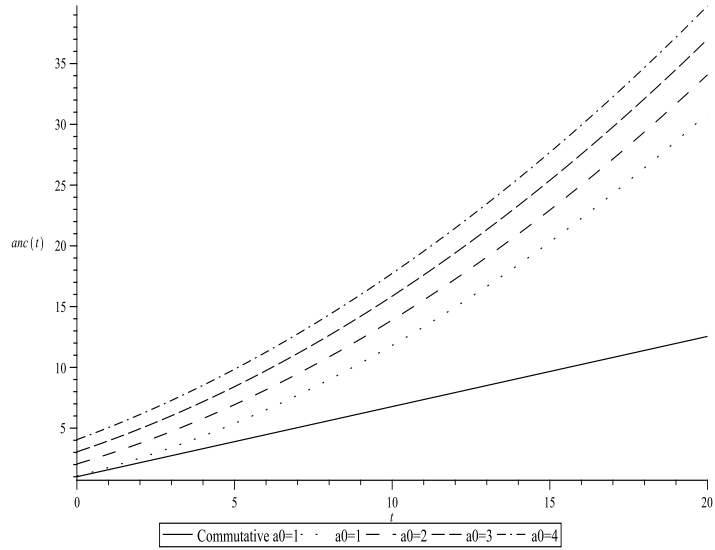}
\end{center}
\caption{$a_{nc}(t)$ as a function of $t$ for different values of $a_0$ and $C = 1$, $\sigma = -0.1$, $\gamma = 0.1$, $\alpha = -0.5$}
\label{k0_a0}
\end{figure}

As we can see from $a_{nc}(t)$ Eq. (\ref{19.0}), if $\alpha \leq 0$, for any positive value of $a_0$ the solutions are always
expansive. If we fix the values of the other parameters and initial conditions, the solutions expand more quickly for greater values of $a_0$ and the initial value of $a_{nc}(t)$ increases when $a_0$ increases. As an example, we may see Figure \ref{k0_a0}.

\subsubsection{Varying $v_0$}

\begin{figure}[!htb]
	\centering
	\begin{minipage}[c]{0.49\linewidth}
		\centering
		\includegraphics[height=7.5cm,width=9.5cm]{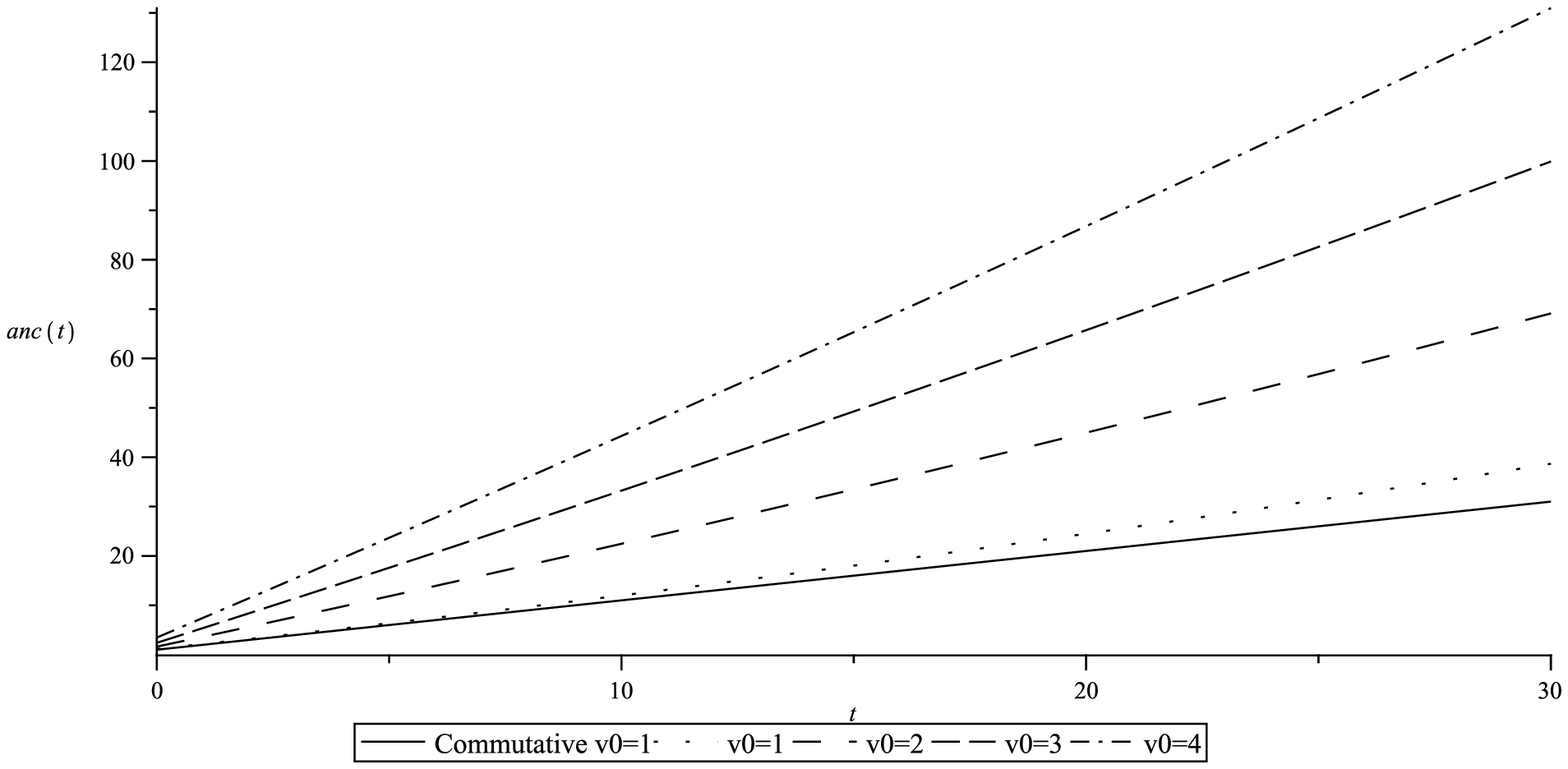}
	\end{minipage}
	\hfill
	\begin{minipage}[c]{0.49\linewidth}
		\centering
		\begin{tabular}{cc}\hline
			$v_0$ & $C$ \\ \hline
			1 & 3.2 \\		
			2 & 12.5 \\
			3 & 27.8 \\
			4 & 49.1 \\ \hline
		\end{tabular}
	\end{minipage}
	\caption{ $a_{nc}(t)$ as a function of $t$ for different values of $v_0$ and $\sigma = -0.1$, $\gamma = 0.1$, $\alpha = -0.1$}
	\label{k0_v0}
\end{figure}

In order to investigate the behavior of $a_{nc}(t)$ Eq. (\ref{19.0}), with respect to the variation of the initial condition
$v_0$, we must use the Friedmann equation (\ref{18.2}). We choose different values of $v_0 \equiv \dot{a}_{c}(t=0)$ and fix all other quantities in Eq. (\ref{18.2}), with the exception of $C$. Therefore, for each value of $v_0$ we choose, we obtain a corresponding value of $C$. Doing that for several different values of $v_0$, we notice that if we increase the values of $v_0$, $C$ also increases. Then, keeping fix the other parameters and initial conditions, $a_{nc}(t)$ Eq. (\ref{19.0}) expands more quickly with the increase of $v_0$. Since, for greater values of $v_0$, we have, also, greater values of $C$, the initial value of $a_{nc}(t)$ increases when $v_0$ increases. As an example, we may see Figure \ref{k0_v0}.

From the above results, for the present case where $k =0$, we notice that $a_{nc}(t)$ Eq. (\ref{19.0}) expands quicker than
$a(t)$ Eq. (\ref{19.1}). The reason for the different rates of expansion between the two scale factors is mostly due to the 
presence of the NC parameter $\alpha$, which has to be negative in order to produce expansive models. In this sense, $\alpha$
may be considered the most important NC parameter, in the present case. Another important property of $a_{nc}(t)$ is that it 
has three free NC parameters. They may be adjusted in order to better describe the data coming from observations.

\subsection{The case k=-1}
\label{k=-1}

Now fixing $k=-1$, the spatial sections have constant negative curvatures. Introducing $k=-1$ in the system Eqs. (\ref{14.8}) and (\ref{14.9}) and in the NC Friedmann equation (\ref{14.95}), we obtain,
\ba
\label{17}
\ddot{a}_{c} & = & a_{c}+\frac{\gamma T_{c}}{2}-\frac{\sigma C}{2}-\frac{\alpha}{6},\\
\label{17.1}
\dot{T_{c}} & = & \frac{\chi}{2}\dot{a_{c}} - 3\sigma a_{c},
\ea
\begin{equation}
\label{17.2}
-3 \dot{a}_{c}^{2}-3 \gamma \dot{a}_{c}+3 a_{c}^{2}+6 a_{c}\left(\frac{\gamma T_{c}}{2}-\frac{\sigma C}{2}\right)+C-\alpha a_{c} = 0.
\end{equation}
After solving, numerically, the system Eqs. (\ref{17}) and (\ref{17.1}), for many different
values of all parameters and initial conditions, the last ones satisfying the NC Friedmann 
equation (\ref{17.2}), we reach the following conclusions.
The general behavior of $a_{nc}(t)$ Eq. (\ref{14.7}) describes a universe that starts
to expand, in an accelerated rate, from its initial size at $t=0$, and continues to expand to an infinity size, after
an infinite time interval. That general behavior of $a_{nc}(t)$ is qualitatively similar to the corresponding commutative
scale factor, the differences being of quantitative nature. 
Now, we want to investigate the behavior of $a_{nc}(t)$ with respect to: $C$, $\alpha$, $\gamma$, $\sigma$, $\chi$, $a_0$, $v_0$ and $T_0$. It is important to remember that, the values of $a_0$ and $T_0$ will be fixed following Eqs. (\ref{ic1}) and (\ref{ic2}), when we are not studying them. When we are studying a given parameter or initial condition, with the exception of $v_0$, we varies that quantity and obtain the value of $v_0$, from the NC Friedmann equation (\ref{17.2}).
When we are studying $v_0$, we choose different values of that initial condition and let $C$ varies. The different values of $C$ are obtained from the NC Friedmann equation (\ref{17.2}). Finally, we want to compare the NC scale factor with the commutative one. In all the examples, we give in the next Subsections, the values of the parameters
and initial conditions are chosen for a better visualization of the results.

\subsubsection{Varying $\alpha$}

\begin{figure}
\begin{center}
\includegraphics[height=7.5cm,width=9.5cm]{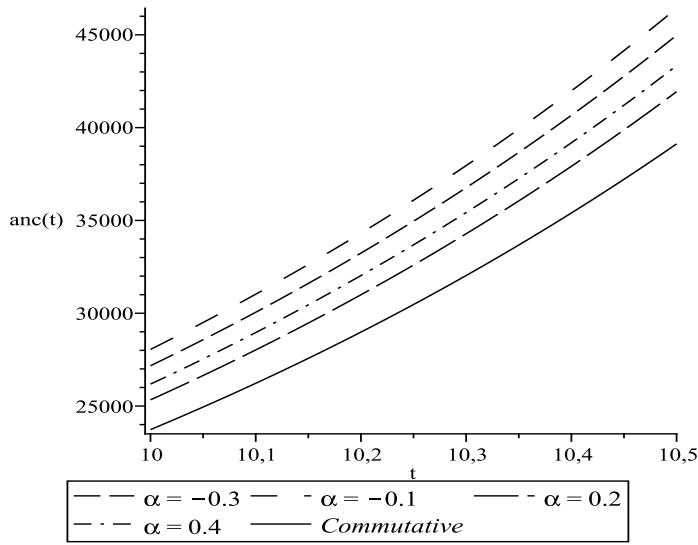}
\end{center}
\caption{$a_{nc}(t)$ as a function of $t$ for different values of $\alpha$ and $C = 1$, $\sigma = -0.01$, $\gamma = 0.4$, 
$\chi = 0.1$}
\label{km1_alpha}
\end{figure}

After solving, numerically, the system Eqs. (\ref{17}) and (\ref{17.1}), for many different values of $\alpha$, 
keeping fix the appropriate initial conditions, we reach the following conclusions.
In the present case, where $k=-1$, a positive or negative NC parameter $\alpha$ gives rise to an expansive $a_{nc}(t)$. Here,
the behavior of the NC scale factor with $\alpha$ is more complicated than in the case where $k=0$. If $\gamma \neq 0$ there is not a simple relationship between the increase or decrease of the NC scale factor expansion rate and the values of $\alpha$. On
the other hand, if $\gamma = 0$, $\sigma \leq 0$ and $\chi$ is positive, negative or zero, the scale factor expands more rapidly for smaller values of $\alpha$. The commutative ($\alpha = \gamma =\sigma =\chi = 0$) solution may expand more rapidly or more slowly than the NC one, depending on the values of the NC parameters. As an example, we may see Figure \ref{km1_alpha}.

\subsubsection{Varying $\gamma$}

\begin{figure}
\begin{center}
\includegraphics[height=7.5cm,width=9.5cm]{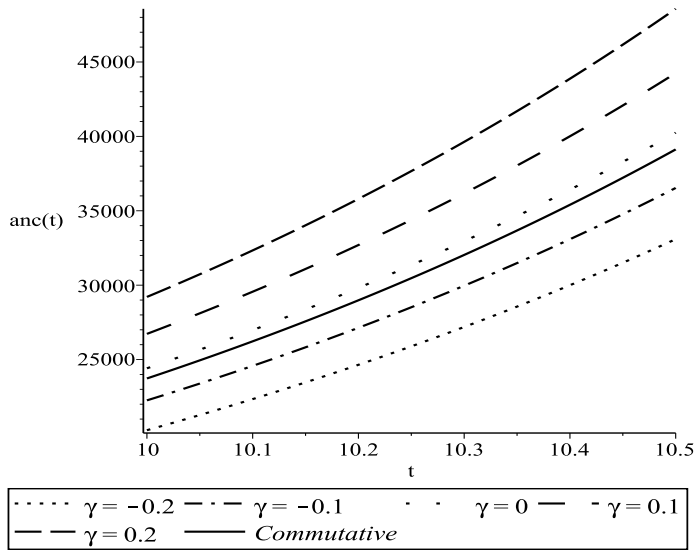}
\end{center}
\caption{$a_{nc}(t)$ as a function of $t$ for different values of $\gamma$ and $C = 1$, $\sigma = -0.1$, $\alpha = 0.1$, 
$\chi = 0.1$}
\label{km1_gamma}
\end{figure}

After solving, numerically, the system Eqs. (\ref{17}) and (\ref{17.1}), for many different values of $\gamma$, keeping fix the appropriate initial conditions, we reach the following conclusions.
Firstly, as in the case where $k=0$, if $T_0 \geq 0$ and $\gamma < 0$, we notice that $a_{nc}(t)$ Eq. (\ref{14.7}) may be negative. Therefore, since $\gamma$ may be positive or negative, in general, we shall restrict our attention to the situation where $T_0 = 0$. It means that, we shall not investigate how $a_{nc}(t)$ depends on $T_0$, for the present models with $k=-1$.
Here, the behavior of the NC scale factor with $\gamma$ is more complicated than in the case where $k=0$. If $\sigma < 0$, $\alpha$ and $\chi$ are positive, negative or zero, the scale factor expands more rapidly for greater values of $\gamma$. Now, if $\sigma = 0$ there are two different options depending the value of
$\chi$: (i) If $\sigma = 0$, $\chi < 0$ and $\alpha$ is positive, negative or zero, the scale factor expands more rapidly for smaller values of $\gamma$; (ii) If $\sigma = 0$, $\chi \geq 0$ and $\alpha$ is positive, negative or zero, the scale factor expands more rapidly for greater values of $\gamma$. Here, as it will be explained below, we shall restrict our attention to the situations where $\sigma \leq 0$. The commutative ($\alpha = \gamma =\sigma =\chi = 0$) solution may expand more rapidly or more slowly than the NC one, depending on the values of the NC parameters. 
As an example, we may see Figure \ref{km1_gamma}.

\subsubsection{Varying $\chi$}

\begin{figure}
\begin{center}
\includegraphics[height=7.5cm,width=9.5cm]{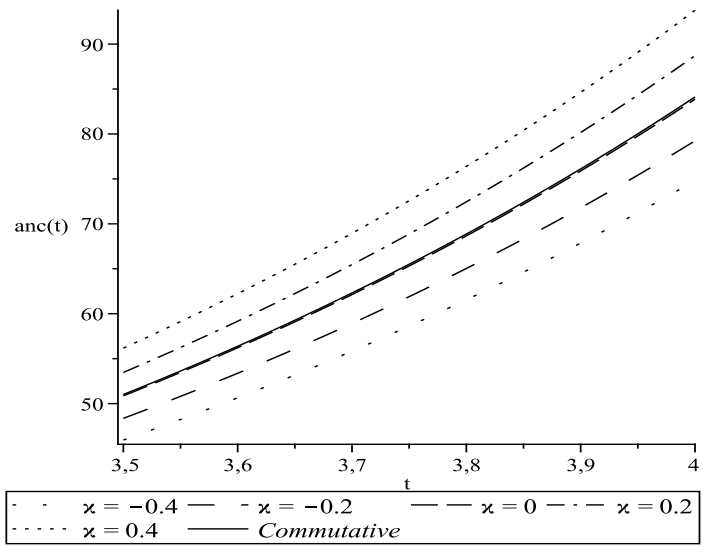}
\end{center}
\caption{$a_{nc}(t)$ as a function of $t$ for different values of $\chi$ and $C = 10$, $\sigma = 0$, $\alpha = 0.1$, 
$\gamma = 0.5$}
\label{km1_chi}
\end{figure}

After solving, numerically, the system Eqs. (\ref{17}) and (\ref{17.1}), for many different values of $\chi$, keeping fix the appropriate initial conditions, we reach the following conclusions.
The NC parameter $\chi$ may be positive, negative or zero. The behavior of $a_{nc}(t)$ as a function of $\chi$ depends on the value of $\gamma$.
There are three different options: 
(i) If $\gamma > 0$, $\alpha$ is positive, negative or zero and $\sigma \leq 0$, the scale factor expands more rapidly for greater values of $\chi$;
(ii) If $\gamma < 0$, $\alpha$ is positive, negative or zero and $\sigma \leq 0$, the scale factor expands more rapidly for smaller values of $\chi$;
(iii) If $\gamma = 0$, $\alpha$ is positive, negative or zero and $\sigma \leq 0$, for all values of $\chi$ the $a_{nc}(t)$ has always the same evolution.
The commutative ($\alpha = \gamma =\sigma =\chi = 0$) solution may expand more rapidly or more slowly than the NC one, depending on the values of the NC parameters.
As an example, we may see Figure \ref{km1_chi}.

\subsubsection{Varying $\sigma$}

\begin{figure}
\begin{center}
\includegraphics[height=7.5cm,width=9.5cm]{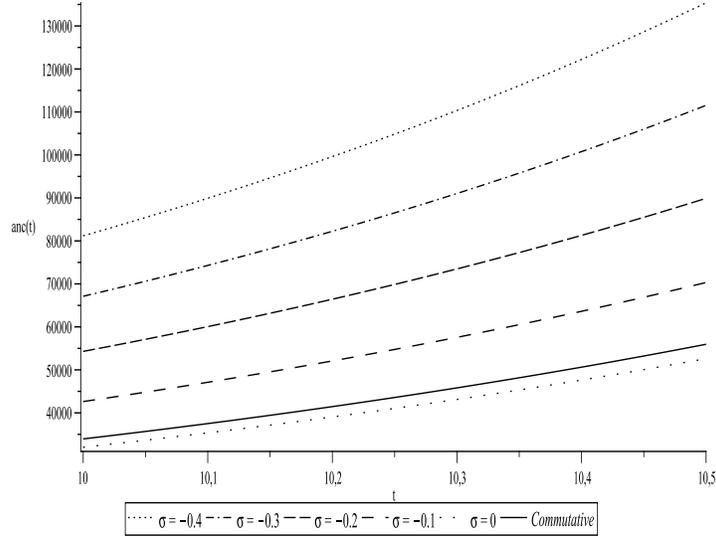}
\end{center}
\caption{$a_{nc}(t)$ as a function of $t$ for different values of $\sigma$ and $C = 10$, $\chi = -0.5$, $\alpha = -0.1$, 
$\gamma = 0.1$}
\label{km1_sigma}
\end{figure}

After solving, numerically, the system Eqs. (\ref{17}) and (\ref{17.1}), for many different values of $\sigma$, keeping fix the appropriate initial conditions, we reach the following conclusions.
As in the case where $k=0$, we shall restrict our attention to $\sigma \leq 0$, because $C \geq 0$ and if $\sigma > 0$, $a_{nc}(t)$ may be negative, from Eq. (\ref{14.7}). Here, the behavior of the NC scale factor with $\sigma$ is more complicated than in the case where $k=0$. The behavior of $a_{nc}(t)$ as a function of $\sigma$ depends on the value of $\gamma$.
There are two different options: 
(i) If $\gamma \geq 0$, $\alpha$ and $\chi$ are positive, negative or zero, the scale factor expands more rapidly for smaller values of $\sigma$;
(ii) If $\gamma < 0$, $\alpha$ and $\chi$ are positive, negative or zero, the scale factor expands more rapidly for greater values of $\sigma$, if we let $a_{nc}(t)$ evolve a sufficient amount of time. For smaller
values of $\sigma \leq 0$, and keeping fix the other parameters and appropriate initial conditions, the initial value of $a_{nc}(t)$ increases.
The commutative ($\alpha = \gamma =\sigma =\chi = 0$) solution may expand more rapidly or more slowly than the NC one, depending on the values of the NC parameters.
As an example, we may see Figure \ref{km1_sigma}.

\subsubsection{Varying $C$}

\begin{figure}
\begin{center}
\includegraphics[height=7.5cm,width=9.5cm]{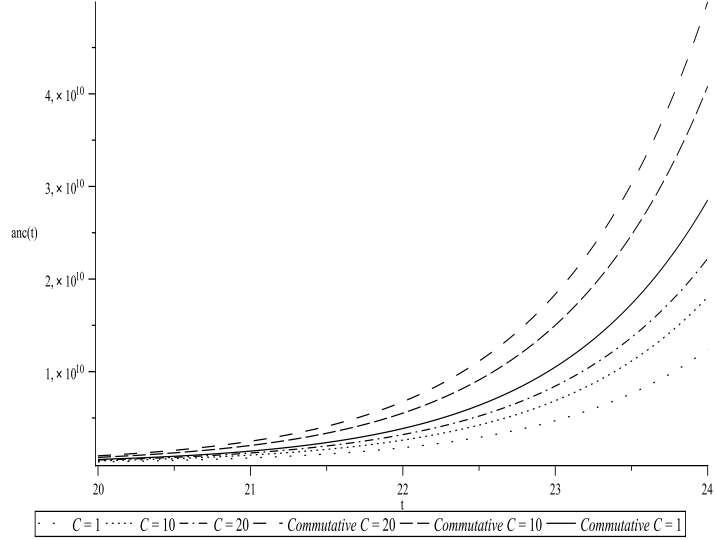}
\end{center}
\caption{$a_{nc}(t)$ as a function of $t$ for different values of $C$ and $\sigma = 0$, $\chi = -0.5$, $\alpha = 0.5$, 
$\gamma = 0.5$}
\label{km1_C}
\end{figure}

After solving, numerically, the system Eqs. (\ref{17}) and (\ref{17.1}), for many different values of $C$, the parameter associated to the fluid energy, keeping fix the appropriate initial conditions, we reach the following conclusions.
For any positive value of $C$ the solutions are always expansive. The NC scale factor expands more quickly for greater values of $C$ and the initial value of $a_{nc}(t)$ increases when $C$ increases. The initial value $v_0$ increases when $C$ increases, from the Friedmann equation (\ref{17.2}). The commutative ($\alpha = \gamma =\sigma =\chi = 0$) solution may expand more rapidly or more slowly than the NC one, depending on the values of the NC parameters. As an example, we may see Figure \ref{km1_C}.

\subsubsection{Varying $a_0$}

\begin{figure}
\begin{center}
\includegraphics[height=7.5cm,width=9.5cm]{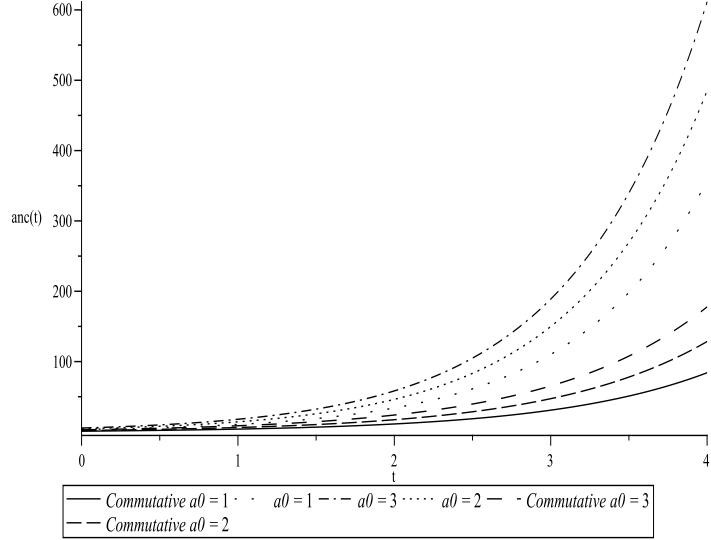}
\end{center}
\caption{$a_{nc}(t)$ as a function of $t$ for different values of $a_0$ and $C=10$, $\sigma = -0.5$, $\chi = 0.5$, $\alpha = 0.5$, $\gamma = 0.5$}
\label{km1_a0}
\end{figure}

After solving, numerically, the system Eqs. (\ref{17}) and (\ref{17.1}), for many different values of $a_0$, keeping fix $T_0$, we reach the following conclusions. If we increase the value of $a_0$, the NC scale factor expands more rapidly. The initial value of $a_{nc}(t)$ increases when $a_0$ increases. The commutative ($\alpha = \gamma =\sigma =\chi = 0$) solution may expand more rapidly or more slowly than the NC one, depending on the values of the NC parameters. As an example, we may see Figure \ref{km1_a0}.

\subsubsection{Varying $v_0$}

\begin{figure}
\begin{center}
\includegraphics[height=7.5cm,width=9.5cm]{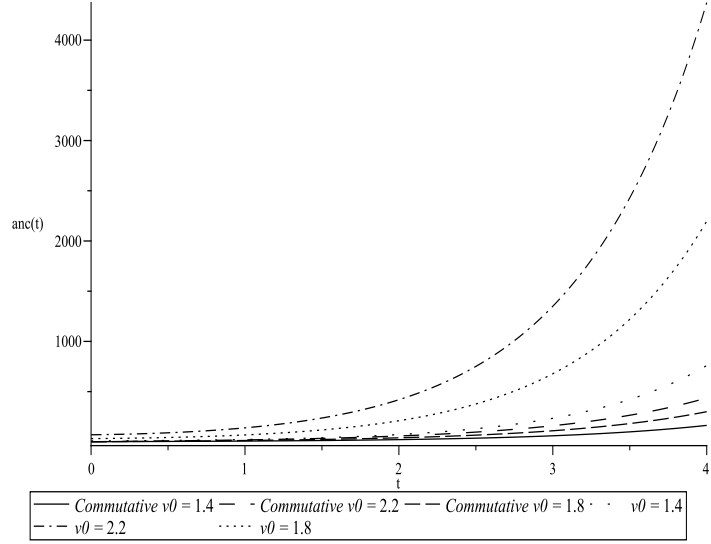}
\end{center}
\caption{$a_{nc}(t)$ as a function of $t$ for different values of $v_0$ and $\sigma = -0.5$, $\chi = 0.5$, $\alpha = -0.5$, $\gamma = 0.5$}
\label{km1_v0}
\end{figure}

After solving, numerically, the system Eqs. (\ref{17}) and (\ref{17.1}), for many different values of $v_0$, keeping fix all other initial conditions, we reach the following conclusions.
If we increase the value of $v_0$, the NC scale factor expands more rapidly.
The initial value of $a_{nc}(t)$ increases when $v_0$ increases. This happens because when we increase
$v_0$, the parameter $C$ also increases and then the initial value of $a_{nc}(t)$ Eq. (\ref{14.7}) increases (remember that $T_0=0$ and
$\sigma<0$).
Since we are letting the parameter $C$ varies, for different values of $v_0$, in the Friedmann equation of
the initial conditions, we notice the following constraint. We must have: $v_0 \geq -\gamma/2 + \sqrt{9\gamma^2 - 
12\alpha + 36}/6$, in order that $C \geq 0$.
The commutative ($\alpha = \gamma =\sigma =\chi = 0$) solution may expand more rapidly or more slowly than the NC one, depending on the values of the NC parameters. As an example, we may see Figure \ref{km1_v0}.

From the above results, for the present case where $k=-1$, we notice that $a_{nc}(t)$ may expand more rapidly or more slowly than the commutative scale factor. The reason for the different rates of expansion between the two scale factors is mostly due to the different values of the NC parameters. Another important property of $a_{nc}(t)$ is that it has four free NC parameters. They may be adjusted in order to better describe the data coming from observations.

\subsection{The case k=1}
\label{k=1}

Let us start by fixing $k=1$, it means that the spatial sections have constant positive curvatures. Introducing $k=1$ in the system Eqs. (\ref{14.8}) and (\ref{14.9}) and in the NC Friedmann equation (\ref{14.95}), we obtain,
\ba
\label{16}
\ddot{a}_{c} & = & -\left(a_{c}+\frac{\gamma T_{c}}{2}-\frac{\sigma C}{2}\right)-\frac{\alpha}{6},\\
\label{16.1}
\dot{T_{c}} & = & \frac{\chi}{2}\dot{a_{c}} + 3\sigma a_{c},
\ea
\begin{equation}
\label{16.2}
-3 \dot{a}_{c}^{2}-3 \gamma \dot{a}_{c}-3 a_{c}^{2}-6 a_{c}\left(\frac{\gamma T_{c}}{2}-\frac{\sigma C}{2}\right)+C-\alpha a_{c} = 0.
\end{equation}
Now we solve, numerically, that system for different values of $\sigma$, $\alpha$, $\chi$, $\gamma$, $C$ and the initial conditions $a_0$, $v_0$ and $T_0$. After solving that system Eqs. (\ref{16}) and (\ref{16.1}), for many different values of all
the parameters and initial conditions, the last ones satisfying the NC Friedmann equation (\ref{16.2}), we found that $a_{nc}(t)$ Eq. (\ref{14.7}) is bounded. It means that it stars from a minimum value at $t=0$, expands up to a maximum value and then contracts to the zero value, after a finite time, given rise to a {\it Big Crunch} singularity. Therefore, since one
of our main motivations, in the present work, is describing the present accelerated expansion of the Universe by NCTY, we shall not investigate, in more details, the present case $k=1$. 

\subsection{Comparison between different values of $k$}
\label{ks}

\begin{figure}
\begin{center}
\includegraphics[height=7.5cm,width=9.5cm]{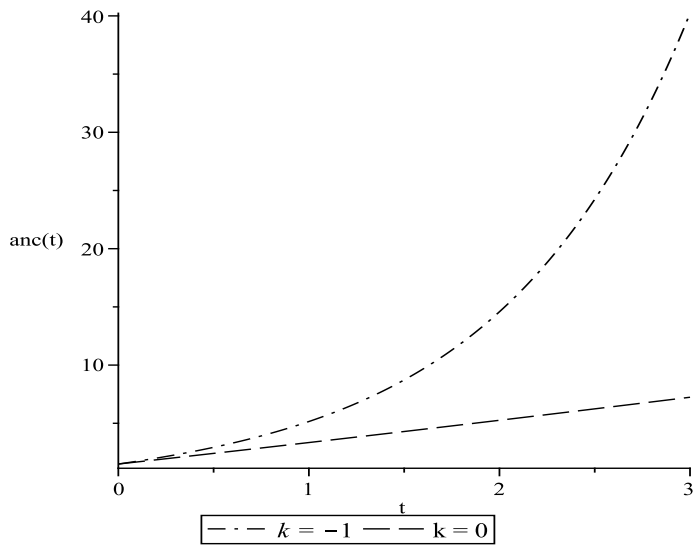}
\end{center}
\caption{$a_{nc}(t)$ as a function of $t$ for different values of $k$ and $C=10$, $\sigma = -0.1$, $\chi = 0.1$, $\alpha = -0.4$, $\gamma = 0.1$}
\label{k0_km1}
\end{figure}

In order to compare the behavior of $a_{nc}(t)$ Eq. (\ref{14.7}), with respect to different values of the curvature parameter
$k$, we restrict our attention to the models where the spatial sections have nil ($k=0$) and negative ($k=-1$) curvatures.
Because only on those models the solutions are expansive. As we have learned in Subsection \ref{k=0}, $a_{nc}(t)$ is expansive
on the models where $k=0$ only if the NC parameter $\alpha \leq 0$. Therefore, we must impose that constraint in order to
compare models with $k=0$ and $k=-1$. Now, we compare $a_{nc}(t)$ Eq. (\ref{19.0}) with the NC scale factor obtained as solution to the system Eqs. (\ref{17}) and (\ref{17.1}), fixing, for both models with $k=0$ and $k=-1$, the same values of the appropriate NC parameters, $C$ and initial conditions. After doing that for many different values of the NC parameters, $C$ and keeping fix $a_0=1$ and $T_0=0$, we reach the following conclusion: for all cases, the $a_{nc}(t)$ for the models with $k=-1$ expand more rapidly than the ones for the models with $k=0$. That result agrees with the corresponding one in the commutative models. As an example of that behavior, we can see Figure \ref{k0_km1}.

\section{Estimates for the values of the NC parameters}
\label{sec:endofuniverse}

In the present section, we want to estimate the values of the NC parameters: $\alpha$, $\gamma$, $\sigma$ and $\chi$. 
Using the present observational data, we can only estimate the values of $\alpha$ and $\gamma$, because, as we shall see, in order to
estimate $\sigma$ and $\chi$, we would need observational data on the variable $T_{nc}(t)$, which are not available. 

In order to obtain estimates for $\alpha$ and $\gamma$, we start with the NC Hamiltonian Eq. (\ref{3,5}), in the gauge $N_{nc} = a_{nc}$.
Let us impose some conditions, in the general NC Hamiltonian, such that it may better describe the present properties of our Universe. First of all, we consider that the spatial sections of the model are flat ($k=0$). Then, we suppose that the matter content of the model consists of
dust ($\omega=0$) and the accelerated expansion is due, entirely, to the presence of the NC parameters. Under those conditions, the NC
Hamiltonian Eq. (\ref{3,5}), reduces to, 
\begin{equation}
\mathcal{H}_{nc}=-\frac{P_{a_{nc}}^{2}}{12}+ P_{T_{nc}}a_{nc}.
\label{25}
\end{equation}
Differently from Sections \ref{sec:general} and \ref{sec:radiation}, here, we are going to work with the NC variables. We shall not introduce
the transformations Eqs. (\ref{5})-(\ref{5.3}). It will show another way to find the NC scale factor as a function of time.

Next, from the above NC Hamiltonian Eq. (\ref{25}), we compute the dynamical equations, with the aid of the deformed Poisson brackets Eqs. 
(\ref{4})-(\ref{4.3}). They are given, to first order in the NC parameters, by,
\ba
\label{movnc1}
\dot{a}_{nc} & = & \left\{a_{nc},H_{nc}\right\} = \frac{-P_{a_{nc}}}{6} + \gamma a_{nc}, \\ 
\label{movnc2}
\dot{P}_{a_{nc}} & = & \left\{P_{a_{nc}},H_{nc}\right\} =  \alpha a_{nc} - P_{T_{nc}}, \\ 
\label{movnc3}
\dot{T}_{nc} & = & \left\{T_{nc},H_{nc}\right\} = \frac{-\chi}{6}P_{a_{nc}} + a_{nc}-\sigma P_{T_{nc}}, \\ 
\label{movnc4}
\dot{P}_{T_{nc}} & = & \left\{P_{T_{nc}},H_{nc}\right\} = \frac{\alpha}{6}P_{a_{nc}} - \gamma P_{T_{nc}}. 
\ea	
As we can see, the NC parameters $\sigma$ and $\chi$ are only present in the dynamical equation for $T_{nc}$ Eq. (\ref{movnc3}). Therefore,
as we have mentioned above, we cannot estimate their values because there is no observational data available on $T_{nc}$.

Now, we want to write a dynamical equation for the NC scale factor $a_{nc}(t)$. Then, we start combining equations (\ref{movnc1}),
(\ref{movnc2}) and (\ref{movnc4}), to find,
\begin{equation}
\label{eq1}
P_{T_{nc}} = C -\alpha a_{nc} + \gamma P_{a_{nc}}.
\end{equation}
If we introduce the above equation (\ref{eq1}) in Eq. (\ref{movnc2}), we may write,
\begin{equation}
\label{eq2}
\dot{P}_{a_{nc}} = 2\alpha a_{nc} - \gamma P_{a_{nc}} - C. 
\end{equation}
From Eq. (\ref{movnc1}), we may take the value of $P_{a_{nc}}$ and introduce it in the above equation (\ref{eq2}), to obtain,
\begin{equation}
\label{eq4}
\dot{P}_{a_{nc}} =  2\alpha a_{nc} +6\gamma\dot{a}_{nc} - C. 
\end{equation}
Finally, the desired equation is obtained after we derive Eq. (\ref{movnc1}) with respect to $t$ and introduce, in the resulting equation, the value of $\dot{P}_{a_{nc}}$ given in Eq. (\ref{eq4}). The dynamical equation for $a_{nc}(t)$ is given by,
\begin{equation}
\label{ancest}
\ddot{a}_{nc} +\frac{\alpha}{3}a_{nc} = \frac{C}{6}. 
\end{equation}
Equation (\ref{ancest}) is a nonhomogeneous, linear, second-order differential equation for $a_{nc}(t)$. It describes an expansive universe if we impose that $\alpha<0$. For an expansive universe, the general solution to that equation has the following form,
\begin{equation}
\label{sol0}
a_{nc}(t)= \left(\frac{a_{0}}{2}+\frac{v_{0}}{2\lambda}+\frac{C}{4|\alpha|}\right)e^{\lambda (t-t_0)} + \left(\frac{a_{0}}{2}-\frac{v_{0}}{2\lambda}+\frac{C}{4|\alpha|}\right)e^{-\lambda (t-t_0)} - \frac{C}{2|\alpha|}, 
\end{equation}
where we explicitly used the fact that $\alpha$ is negative, $\lambda = \sqrt{\frac{|\alpha|}{3}}$, $a_{nc}(t=t_0)\equiv a_0$ and $\dot{a}_{nc}(t=t_0)=v_0$. Since, the matter content of our model is dust, we shall consider $t_0$ and $a_0$ as the time and the scale factor at the epoch of matter-radiation equality. After that epoch, the matter dominates the Universe. Those quantities have the following approximated values: $t_0 \approx 3,5\times10^{5}$ years and $a_0 \approx 2,84\times10^{-4}$ \cite{Liddle.2003}. It is possible to write $v_0$ in terms of $a_0$. In order to do that, we need the NC Friedmann equation for the present model. That equation may be obtained by imposing the constraint that $\mathcal{H}_{nc}$ Eq. (\ref{25}) is zero. With the aid of equations (\ref{movnc1}) and (\ref{eq1}), the NC Friedmann equation is given by,
\be
\label{frdNC}
3\dot{a}_{nc}^{2} - (C + |\alpha|a_{nc})a_{nc} =0.
\ee
Finally, introducing the initial values of $a_{nc}(t)$ and $\dot{a}_{nc}(t)$ in the above equation (\ref{frdNC}), we obtain,
\be
\label{v0}
v_0 = \sqrt{\frac{(C + |\alpha|a_0)a_0}{3}},
\ee
where we are considering only positive values of $v_0$, because we want expansive solutions.

Unfortunately, observing $a_{nc}(t)$ Eq. (\ref{sol0}) and the initial condition Eq. (\ref{v0}), we notice that they depend only on $\alpha$,
to first order in the NC parameters. Therefore, we may only estimate the value of $\alpha$, from the present model. In order to do that, we shall suppose that the present accelerated expansion of the Universe has started at a certain time ($t^{ae}$) and scale 
factor ($a_{nc}^{ae}$). The ten values of those quantities, we used in order to obtain ten estimated values of $\alpha$, 
are shown in the first two columns of Table \ref{comparison}. The other important quantity, we must furnish is the constant $C$,
associated to the dust density energy. That quantity may be written as: $C=\Omega_{mat}H^{2}$, where $\Omega_{mat}$ is the matter 
density parameter and $H$ is the Hubble constant. We shall suppose that $C$ may be well described by the present values of the
matter density parameter and the Hubble constant: $C=\Omega_{mat0}H_{0}^{2}$ where $\Omega_{0}=0,2825$ and $H_{0}=70(km/s)/Mpc$ 
\cite{riess0} and \cite{perlmutter}. Finally, we numerically solve equation (\ref{sol0}) with the aid of Eq. (\ref{v0}) and find the
ten values of $\alpha$, given in the third column of Table \ref{comparison}. The last column of that Table is the numerical error
obtained by using the values of $\alpha$, calculated, in order to compute $a_{nc}^{ae}$, from equation (\ref{sol0}) with the aid of 
Eq. (\ref{v0}). Observing the values of $\alpha$ from Table \ref{comparison}, we observe, initially, that they are very small. Then, we notice
that $\alpha$ increases as $t^{ae}$ approaches the initial moments of the Universe. That result is what one expects since noncommutativity must had been more important at the beginning of the Universe. 

\begin{table}[h!]
\caption{{\protect\footnotesize {Table with $10$ different estimated values of $\alpha$.
}}}
\centering
{\scriptsize\begin{tabular}{|c|c|c|c|}
\hline $a_{nc}^{ae}$ & $t^{ae}$ ($10^9$ years) & $\alpha$ & numerical error (\%)\\ \hline
$1.0$ & $13.4560$ & $-7.776000000\times10^{-34}$ & 0.0114869\\ \hline
$0.9$ & $12.0224$ & $-9.874000000\times10^{-34}$ & 0.00253033\\ \hline
$0.8$ & $10.5167$ & $-1.309000000\times10^{-33}$ & 0.01502281\\ \hline
$0.7$ & $8.9511$ & $-1.832000000\times10^{-33}$ & 0.00751010\\ \hline
$0.6$ & $7.3488$ & $-2.746000000\times10^{-33}$ & 0.04690716\\ \hline
$0.5$ & $5.7470$ & $-4.500000000\times10^{-33}$ & 0.03459550\\ \hline
$0.4$ & $4.1973$ & $-8.320000000\times10^{-33}$ & 0.0145302\\ \hline
$0.3$ & $2.7629$ & $-1.839000000\times10^{-32}$ & 0.07414757\\ \hline
$0.2$ & $1.5148$ & $-5.570000000\times10^{-32}$ & 0.0393519\\ \hline
$0.1$ & $0.5370$ & $-3.592000000\times10^{-31}$ & 0.0059710\\ \hline
\end{tabular}
}
\label{comparison}
\end{table}

\section{Conclusions}
\label{sec:conclusions}

In the present work, we have tried to explain the present accelerated expansion of our Universe by means of a small, residual, NCTY
between the geometrical and matter phase space variables of a FRW cosmological model ($a$, $P_a$, $T$, $P_T$). The matter content of
the model can be any perfect fluid with an equation of state of the form Eq. (\ref{2.0}). We considered the most general situation where all 
the Poisson brackets between the variables and the canonically conjugated momenta are nonzero. That procedure gives rise to four NC 
parameters ($\alpha$, $\gamma$, $\sigma$, $\chi$), which greatly modifies the scale factor dynamics of the corresponding commutative 
model. We explicitly saw it in Section \ref{sec:radiation}, where we considered a FRW cosmological model where the matter content is a radiation perfect fluid and the constant curvature of the spatial sections could be positive, negative or nil. There, we studied 
how each NC parameter: $\alpha$, $\gamma$, $\sigma$, $\chi$, modifies the corresponding commutative cosmological model. In particular, we
learned that only for models where the constant curvature of the spatial sections are negative or nil, the scale factor is expansive. For models where the curvature of the spatial sections are nil, we found an algebraic solution for the system of equations. From that solution 
Eq. (\ref{19.0}), we saw that only three NC parameters ($\alpha$, $\gamma$, $\sigma$), are explicitly present. Studying that solution, we learned that: (i) $\alpha$ is the most important NC parameter and it has to be negative in order to produce expansive solutions; (ii) for smaller values of $\alpha$ and $\gamma$, the universe expands more rapidly; (iii) $\sigma$ gives a constant contribution and has to be 
negative or zero; (iv) if one increases the fluid energy parameter $C$ and the initial conditions $a_0$ and $v_0$, $a_{nc}(t)$ 
Eq. (\ref{19.0}) expands more rapidly; (v) finally, the NC scale factor expands more rapidly than the commutative one. Now, for models where the curvature of the spatial sections are negative, we had to solve the system of equations numerically. From the numerical study, we found that all four NC parameters ($\alpha$, $\gamma$, $\sigma$, $\chi$) contribute to the solution. From the numerical studies, we learned that: (i) $\alpha$ may be negative or positive and if $\gamma = 0$, $\sigma \leq 0$ and $\chi$ is positive, negative or zero, the NC scale factor expands more rapidly for smaller values of $\alpha$; (ii) the behavior of the NC scale factor with $\gamma$ is more complicated than in the case where $k=0$, it depends mainly on the value of $\sigma$; (iii)  the behavior of the NC scale factor with $\sigma$ is more complicated than in the case where $k=0$, it depends mainly on the value of $\gamma$. $\sigma$ has to be negative or zero; (iv) the behavior of $a_{nc}(t)$ as a function of $\chi$ depends on the value of $\gamma$; (v) if one increases the fluid energy parameter $C$ and the initial conditions $a_0$ and $v_0$, $a_{nc}(t)$ expands more rapidly; (vi) finally, the NC scale factor may expand more rapidly or more slowly than the commutative
scale factor, depending on the values of the NC parameters. When one compares the two models with $k=0$ and $k=-1$, one notices that the NC
scale factor expands more rapidly for the models with $k=-1$.

In Section \ref{sec:endofuniverse}, we were able to estimate the value of the NC parameter $\alpha$ studying a FRW model, with $k=0$ and a dust
perfect fluid. The estimated values of $\alpha$ are very small and they increase as $t^{ae}$ approaches the initial moments of the Universe. That result is what one expects since noncommutativity must had been more important at the beginning of the Universe. As we saw here, the NC models can produce an accelerated expansion of the universe. Another important property of those models is the fact that they have additional free parameters, with respect to the corresponding commutative models. In the present models, for a generic perfect fluid, four additional NC parameters, which may be used to better adjust the observational data.

{\bf Acknowledgements}. This study was financed in part by Funda\c{c}\~{a}o de Amparo a Pesquisa de Minas Gerais (FAPEMIG) and Coordena\c{c}\~{a}o de Aperfei\c{c}oamento de Pessoal de N\'{i}vel Superior (CAPES) - Finance code 001.

\end{document}